\documentclass[english]{revtex4}
\usepackage[T1]{fontenc}
\usepackage[latin9]{inputenc}
\usepackage{amssymb}

\usepackage{babel}

\begin{document}

\title{Constraint on the Mass of Primordial Black Holes from the Cosmological
Constant}

\author{A. Larrañaga}

\email{ealarranaga@unal.edu.co}

\affiliation{Universidad Nacional de Colombia. Observatorio Astronómico Nacional
(OAN)}

\affiliation{Universidad Distrital Francisco Jose de Caldas. Facultad de Ingeniería.}

\author{J. M. Tejeiro }

\email{jmtejeiros@unal.edu.co}

\affiliation{Universidad Nacional de Colombia. Observatorio Astronómico Nacional
(OAN)}
\begin{abstract}
In a recent propposal, the cosmological constant has been considered
as as a new thermodynamical variable and its change is related to
the radiation produced by black holes. Using this consideration and
by modelling the primordial black holes as Schwarzschild-de Sitter
holes,we have constrained the total mass of primordial black holes
evaporated by now, giving an estimate of the order of $1.624\times10^{24}M_{\odot}$. 
\end{abstract}
\maketitle
In the early universe, the great compression associated with the Big
Bang could have formed black holes with different masses. These are
known as Primordial Black Holes (PBH) and are an important because
they could provide some proofs about these early stages of the universe,
the gravitational collapse, the quantum evaporation and about quantum
gravity. The number of PBH is unknown but some estimated densities
have been calculated \cite{carr}. In particular, PBH constraints
can be obtained by considering a varying gravitational ''constant''
$G$, that can be different at early times. The simplest example is
the Brans-Dicke theory, in which $G$ is associated with a scalar
field that can be non-homogeneus. Therefore, the cosmological consequences
of the variation of $G$ depend on the PBH evaporation and how evolves
the value of $G$ near the black hole\cite{barrow1,barrow}. 

On the other hand, the cosmological constant has an important role
in the evolution of the universe, including the early universe and
the Big Bang. Recently, some authors claimed that the cosmological
constant can be promoted to a thermodynamical state variable \cite{sekiwa,wang}.
Then, the first law in differential form, is modified to be

\begin{equation}
dM=TdS+\Omega dJ+\Phi dQ+\Theta d\Lambda,\end{equation}
where $\Theta$ is the generalized volume conjugate to the cosmological
constant $\Lambda$. The integral mass formula is generalized to

\begin{equation}
\frac{n-3}{n-2}M=TS+\Omega J+\frac{n-3}{n-2}\Phi Q+\frac{1}{n-2}\Theta\Lambda,\label{eq:intfirst}\end{equation}
where $n$ is the dimension of the spacetime. Using this interpretation,
it is argued that the decrease of vacuum energy, represented by the
cosmological constant, is equal to the decreasing of entropy inside
the black hole horizon. Therefore, for an external observer, this
process is seen as if the vacuum energy is transformed quantum mechanically
to the energy of radiation of the black hole. This effect attributes
the Hawking radiation to the varying cosmological constant.

In this paper, we will consider the inverse process, i.e. that the
Hawking radiation produced by the evaporation of a black hole may
produce a variable cosmological constant. In particular, we will consider
the observational value of the cosmological constant today to constrain
the mass of the PBH produced in the early universe using this process.

\section{Primordial Black Holes}

In the early universe, black holes with a wide range of masses could
have formed as a consecuence of the compression associated with the
Big Bang. These objects are known as Primordial Black Holes (PBH)
and their horizon mass $M$ depends on their formation epoch. If $t$
is the time after the Big Bang, the comparison between the cosmological
density and the density of a black hole shows that the mass of a PBH
formed at time $t$ is given by\cite{carr}

\begin{equation}
M\left(t\right)\approx\frac{c^{3}t}{G}\approx10^{15}\left(\frac{t}{10^{-23}s}\right)g.\label{eq:PBHmass}\end{equation}

At early times, the mass of a PBH is really small. For example, at
Planck's time, $t=10^{-43}s$, the mass of a PBH is just $M=10^{-5}g$.
These small masses indicate that thermodynamical effects need to be
considered because, as is well known, the Hawking temperature depends
on the black hole's mass. Therefore, a PBH has a temperature of the
order

\begin{equation}
T=\frac{\hbar c^{3}}{8\pi Gk_{B}M}\approx10^{-7}\left(\frac{M}{M_{\odot}}\right)^{-1}K,\end{equation}
i.e. that the smaller the mass of the black hole, the higher the temperature.
Since the energy of the radiation is supplied by the black hole's
mass, it is expected that, after some time, the black hole evaporates.
The time scale of this process of evaporation depends on $M$. For
a PBH, it is given by

\begin{equation}
\tau\left(M\right)\approx\frac{\hbar c^{4}}{G^{2}M^{3}}\approx10^{64}\left(\frac{M}{M_{\odot}}\right)^{3}\mbox{years}.\end{equation}
This timescale shows that black holes with masses smaller than $10^{15}g$
have evaporated completely by now and equation (\ref{eq:PBHmass})
implies that these PBH formed before $t=10^{-23}s$.

\section{Schwarzschild-de Sitter Black Hole}

The Scharzschild-de Sitter (SdS) space time in its static form is
given by the line element

\begin{equation}
ds^{2}=f\left(r\right)dt^{2}-\frac{1}{f\left(r\right)}dr^{2}-r^{2}d\Omega^{2}\end{equation}
 where

\begin{equation}
f\left(r\right)=1-\frac{2M}{r}-\frac{\Lambda}{3}r^{2}.\end{equation}
This is a spherically symmetric spacetime with two parameters: $M$,
the mass of the black hole and $\Lambda>0$, the cosmological constant.
The horizons of this solution are defined by the equation

\begin{equation}
1-\frac{2M}{r}-\frac{\Lambda}{3}r^{2}=0,\end{equation}
that corresponds to a third order polynomial. The solutions of this
equation are parameterized by trigonometric functions and their inverses
\cite{arraut}, as 

\begin{eqnarray}
r_{1} & = & -\frac{2}{\sqrt{\Lambda}}\cos\left(\frac{1}{3}\cos^{-1}\left(3M\sqrt{\Lambda}\right)\right)\\
r_{2} & = & -\frac{2}{\sqrt{\Lambda}}\cos\left(\frac{1}{3}\cos^{-1}\left(3M\sqrt{\Lambda}\right)+2\pi\right)\\
r_{3} & = & -\frac{2}{\sqrt{\Lambda}}\cos\left(\frac{1}{3}\cos^{-1}\left(3M\sqrt{\Lambda}\right)+4\pi\right).\end{eqnarray}

The involved trigonometric functions in these solutions forbid any
result with $3M\sqrt{\Lambda}>1$, and gives a maximum mass for the
black hole,

\begin{equation}
M_{max}=\frac{1}{3\sqrt{\Lambda}}.\end{equation}
Note that the first solution $r_{1}$ has a maximum value

\begin{equation}
r_{1}\left(M_{max}\right)=-\frac{2}{\sqrt{\Lambda}},\end{equation}
that shows how this root is always negative. Thus, the SdS black hole
has only two physical horizons $r_{2}$ and $r_{3}$. These roots
can be expanded retaining corrections up to third order in $M\sqrt{\Lambda}$,
\cite{arraut}, as

\begin{equation}
r_{2}\approx\sqrt{\frac{3}{\Lambda}}-M\end{equation}

\begin{equation}
r_{3}\approx2M\left(1+\frac{4}{3}M^{3}\Lambda^{3/2}\right).\end{equation}

When considering the cosmological constant as a new thermodynamical
state variable, the integral form of the first law for this kind of
black hole has the form of equation (\ref{eq:intfirst}) where the
new term, $\Theta\Lambda$, has dimensions of energy. Since the cosmological
constant is related with the vacuum energy density,

\begin{equation}
\rho_{vac}=\frac{\Lambda}{8\pi},\end{equation}
the function $\Theta$ is interpreted as a generalized volume. For
a Schwarzschild-de Sitter black hole, $\Theta$ is given by\cite{sekiwa} 

\begin{equation}
\Theta=\left(\frac{\partial M}{\partial\Lambda}\right)_{S}=-\frac{r_{H}^{3}}{6}.\end{equation}
where $r_{H}$ is the radius of the event horizon. As can be seen,
$\Theta$ corresponds to the volume of the region occupied by the
black hole, with a pre-factor. Thus, the thermodynamical contribution
$\Theta\Lambda$ can be written as\begin{equation}
\Theta\Lambda=-\frac{4\pi r_{H}^{3}}{3}\left(\frac{\Lambda}{8\pi}\right),\end{equation}
that is exactly the product of the vacuum energy density and the volume
inside the event horizon of the black hole. Therefore, the first law
of thermodynamics is now

\begin{equation}
dM=TdS+\Theta d\Lambda.\label{eq:SdSfirstlaw}\end{equation}

\section{Constraint on the PBH Mass by the Cosmological Constant Value}

If we consider that PBH can be modelled as Schwarzschild-de Sitter
black holes, and we suppose that their evaporation process does not
affect the entropy, i.e. $dS=0$, the firts law (\ref{eq:SdSfirstlaw})
gives

\begin{equation}
dM=\Theta d\Lambda=-\frac{4\pi r_{H}^{3}}{3}d\left(\frac{\Lambda}{8\pi}\right),\end{equation}
that can be interpreted by saying that the decrease of mass by Hawking
evaporation produces a change of the cosmological constant. The ubication
of the event horizon for the Schwarzschild-de Sitter black hole is
given as the larger of the roots $r_{2}$ and $r_{3}$. By comparing
the observational value of the cosmological constant and the maximum
value of the masses of PBH that have evaporated by now, we conclude
that the event horizon corresponds to the $r_{2}$ root, that is approximated
by

\begin{equation}
r_{H}=r_{2}\approx\sqrt{\frac{3}{\Lambda}}.\end{equation}
Thus, the first law becomes

\begin{equation}
dM=-4\pi\sqrt{3}\Lambda^{-3/2}d\left(\frac{\Lambda}{8\pi}\right).\end{equation}

This equation can be integrated to obtain

\[
M-M_{0}=\Delta M_{PBH}=\sqrt{\frac{3}{8\pi}}\sqrt{8\pi}\Lambda^{-1/2},\]
where $M_{0}$ is an integration constant that can be interpreted
as the initial mass of the black hole. Then, the vacuum energy density
can be expresed as

\begin{equation}
\rho_{vac}=\frac{\Lambda}{8\pi}=\frac{3}{8\pi\left(\Delta M_{PBH}\right)^{2}},\end{equation}
where $\Delta M_{PBH}$ represents the mass of primordial black holes
that is completely evaporated at some epoch. Recent cosmological observations
imply that the cosmological constant has the limit value\cite{carroll} 

\begin{equation}
\left|\rho_{vac}^{obs}\right|\leq2\times10^{-10}\frac{erg}{cm^{3}}.\end{equation}

Therefore, the estimate of $\rho_{vac}$ constrains the PBH mass that
is completely evaporated by now to

\begin{equation}
\Delta M_{PBH}\geq\left[\frac{1}{144\pi}\times10^{10}\frac{cm^{3}}{erg}\right]^{1/2}\end{equation}

\begin{equation}
\Delta M_{PBH}\geq1.148\times10^{57}gr.\end{equation}

In solar masses, this result is,approximately,

\begin{equation}
\Delta M_{PBH}\gtrsim1.624\times10^{24}M_{\odot}.\end{equation}

\section{Conclusion}

When considering the cosmological constant as a new thermodynamical
variable, its change is related to the radiation produced by black
holes. By modelling the primordial black holes as Schwarzschild-de
Sitter holes and using the observational value of the cosmological
constant today we have constrained the total mass of primordial black
holes evaporated by now to be of the order of $1.624\times10^{24}M_{\odot}$.

\end{document}